%Paper: gr-qc/9501018
%From: Jose M Izquierdo <J.M.Izquierdo@damtp.cambridge.ac.uk>
%Date: Tue, 17 Jan 1995 17:24:08 GMT

\input phyzzx

%%%%%%%%%%%%%%%%%%%%%%%%%%%%%%%%%References%%%%%%%%%%%%%%%%%%%%%%%%%%%%%%%%%%%

\REF\AT {A. Ach{\'u}carro and P.K. Townsend, Phys. Lett. {\bf 180B} (1986) 89;
A. Ach{\'u}carro, MSc. Thesis, Univ. of the Basque country, Bilbao, (1986).}
\REF\BTZ {M. Ba{\~ n}ados, C. Teitelboim, and J. Zanelli, Phys. Rev.
Lett.  {\bf 69}
(1992) 1849; M. Ba{\~ n}ados, M. Henneaux, C. Teitelboim and J.
Zanelli, Phys.  Rev.
{\bf D48} (1993) 1506.}
\REF\HC {O. Coussaert and M. Henneaux, Phys. Rev. Lett. {\bf 72} (1994) 183.}
\REF\DJ {S. Deser and R. Jackiw, Ann. Phys. (N.Y.) 153, (1984) 405.}
\REF\AD {L. Abbott and S. Deser, Nucl. Phys. {\bf B195} (1982) 76.}
\REF\AW {E. Witten, Nucl. Phys. {\bf B311} (1988/1989) 46.}
\REF\MS {P. Menotti and D. Seminara, {\it Energy theorem for 2+1
dimensional  gravity}, preprint MIT-CTP 2324, IFUP-TH-33/94.}
\REF\GH {G.W. Gibbons and C.M. Hull, Phys. Lett. {\bf 109B} (1982) 190.}
\REF\GKLTT {G.W. Gibbons, D. Kastor, L. London, P.K. Townsend and J.
Traschen,  Nucl. Phys. {\bf B}, Nucl. Phys. {\bf B416} (1994) 850.}
\REF\KK {E. Witten, Nucl. Phys. {\bf B195} (1982) 481.}
\REF\RM {R.B. Mann and S.F. Ross, Phys. Rev. D{\bf 47} (1993) 3319; Y.
Peleg  and A. Steif, {\it
Phase transition for gravitationally collapsing dust shells in 2+1
dimensions}, preprint
WISC-MILW-94-TH-26/UCD-PHY-94-40.}
\REF\dWNT {B. de Wit, H. Nicolai and A.K. Tollsten, Nucl. Phys. {\bf
B392}  (1993) 3.}
\REF\GGRS {S.J. Gates, Jr., M.T. Grisaru, M. Ro{\v c}ek and W. Siegel, {\it
Superspace, or One Thousand and One Lessons in Supersymmetry},
(Benjamin/Cummings 1983).}
\REF\SSS{A.M. Perelomov, Phys. Rep.  {\bf 174} (1989) 229.}
\REF\Witone{ E. Witten, Commun. Math. Phys. 80 (1981) 381; J. Nester,
Phys.  Lett. {\bf A83}
(1981) 241.} \REF\GHW{G.W. Gibbons, C.M. Hull and N. Warner, Nucl.
Phys. {\bf  B218} (1983) 545.}
\REF\WZ {F. Wilczek and A. Zee, Phys. Rev. Lett. {\bf 51} (1983) 2250.}
\REF\BH {J.D. Brown and M. Henneaux, Commun. Math. Phys. {\bf 104} (1986) 207.}
\REF\HW {G.T. Horowitz and D. Welch, Phys. Rev. Lett. {\bf 71} (1993) 328.}

%%%%%%%%%%%%%%%%%%%%%%%%%%%%%%%%%%%%Macros%%%%%%%%%%%%%%%%%%%%%%%%%%%%%%%%%%%%
%%%%%%%%%%%%%%%%%%%%%%%%%%%%%%%%%%%%%%%%%%%

\def\tr{{\rm tr}}

\font\mybb=msbm10 at 12pt
\def\bb#1{\hbox{\mybb#1}}
\def\R {\bb{R}}

\def\uzero{{\underline 0}}
\def\uone{{\underline 1}}
\def\utwo{{\underline 2}}

\def\square{\kern1pt\vbox{\hrule height 0.5pt\hbox
{\vrule width 0.5pt\hskip 2pt
\vbox{\vskip 4pt}\hskip 2pt\vrule width 0.5pt}\hrule height
0.5pt}\kern1pt}

\def\cM{{\cal M}}

\def\dslash{\partial\hskip-6pt\slash^{\phantom 8}\!}

\def\dtwoslash{{}^{(2)}{\cal D}\hskip-8pt\slash^{\phantom 8}}
\font\bn=msam10 scaled 1200

\def\geq{\hbox{\bn\char '76}}

%%%%%%%%%%%%%%%%%%%%%%%%%%%%%%%%%%%%TITLE%%%%%%%%%%%%%%%%%%%%%%%%%%%%%%%%%%%%
%%%%%%%%%%%%%%%%%%%%%%%%%%%%%%%%%%%%PAGE%%%%%%%%%%%%%%%%%%%%%%%%%%%%%%%%%%%%%%
%%%%%%%%%%%%%%%%%%%%%%%%%%%%%%%%%%%%%%%%%

\Pubnum{ \vbox{ \hbox{R/94/44} \hbox{9501018}} }
\pubtype{}
\date{December, 1994}

\titlepage

\title{Supersymmetric spacetimes in (2+1) adS-supergravity models}

\author{J.M. Izquierdo and P.K. Townsend}
\address{DAMTP, Univ. of Cambridge, Silver St., Cambridge, U.K. }

\abstract {We find a class of (2+1)-dimensional spacetimes admitting
Killing  spinors appropriate to (2,0) adS-supergravity. The vacuum
spacetimes  include anti-de Sitter (adS)  space and charged extreme
black  holes, but there are many others, including spacetimes of
arbitrarily large negative energy that have only conical singularities, and the
spacetimes of  fractionally  charged point particles. The non-vacuum
spacetimes are those of self-gravitating  solitons obtained
by coupling (2,0)  adS supergravity to sigma-model matter. We show,
subject to  a condition on the matter currents (satisfied by the sigma
model), and a conjecture concerning global obstructions to the
existence of certain types of spinor fields, that  the mass of
each supersymmetric spacetime saturates a classical bound, in terms of
the angular momentum and charge, on the total energy of arbitrary
field  configurations with the same boundary conditions,
although these bounds may  be violated quantum mechanically.}

\endpage

\pagenumber=2

%%%%%%%%%%%%%%%%%%%%%%%%%%%%%%%Chapter1%%%%%%%%%%%%%%%%%%%%%%%%%%%%%%%%

\chapter{ Introduction}

Theories of gravity in 2+1 dimensions continue to be a fertile area
for the investigation of the consequences of general covariance in
field  theory. The  (p,q) anti-de Sitter (adS) supergravity
theories [\AT], which can be viewed as pure Chern-Simons (CS) terms
for the  superalgebras $osp(p|2;\R) \oplus osp(q|2;\R)$, are of
particular  interest for various reasons. Firstly, the
(0,0) case is just 2+1 Einstein gravity with a negative cosmological
constant,  which has been
shown to admit asymptotically-adS black hole solutions [\BTZ]. This
theory is  the bosonic sector
of the (1,1) theory and in this context one can ask whether black hole
solutions are
`supersymmetric' in the sense of preserving some of the supersymmetry
of the  adS vacuum. This is equivalent to asking whether the black
hole  spacetime admits a suitably-defined Killing spinor
and it has been shown [\HC] that only the extreme rotating black
holes, and  the non-rotating
`black hole vacuum' [\BTZ], do so. Secondly, for $p$ or $q$ greater
than one,  the action includes an $so(p)\oplus so(q)$ pure CS term.
The  simplest such case is the (2,0) theory which we review
below. We find a one-function class of `off-shell' bosonic field
configurations of this model that preserve some supersymmetry, i.e.
admit  Killing spinors. The function is fixed by the field
equations and the boundary conditions, and depends on the adS inverse
radius, $m$. For zero charge, $Q$, we find a class of supersymmetric
{\it  vacuum} spacetimes with ADM mass $M= 2mJ-n^2$
(relative to the `black hole vacuum'), and total angular momentum $J$,
for any  integer $n$. When $n=0$ or $|n|=1$ we recover, for $J=0$, the
black-hole vacuum and anti-de Sitter spacetime
respectively. For $|n|>1$ we find a new class of supersymmetric
spacetimes  with a naked conical singularity of negative deficit angle
$\delta= -2\pi(|n|-1)$. The point particle spacetimes [\DJ]
with $-1>M>0$, which have a naked conical singularity of positive
deficit  angle $\delta= 2\pi (1+
\sqrt{M})$, are also supersymmetric  when $Q$ is such that $M= -4Q^2$.

It is a general feature of supergravity theories that solutions that
are  supersymmetric (in the above sense of admitting a Killing spinor)
saturate a Bogomolnyi-Gibbons-Hull type of bound on the ADM energy of
any  field configuration that is either non-singular or, if
singular, can evolve from an initially non-singular configuration. A
special  case of this bound is the positivity of the ADM energy. One
usually  finds (e.g. for asymptotically flat spacetimes
of dimension $d>3$) that, once boundary conditions at spatial infinity
have  been specified, there is a unique supersymmetric zero charge
state which  is naturally identified as the vacuum. An exception to
this rule is (2,0) $(adS)_3$-supergravity because it turns out that
there are many
supersymmetric zero charge configurations. The only non-singular one
is the  adS spacetime itself
but the black hole vacuum, which has a null singularity, is also
supersymmetric and is non-singular on all spacelike hypersurfaces
orthogonal  to the orbits of the timelike Killing
vector field. Other supersymmetric zero charge spacetimes have naked
timelike  conical singularities and do not admit a non-singular Cauchy
surface, so only adS spacetime and the black hole vacuum are
candidates for  `the' vacuum. The reason that there can be {\it two} such
candidates is that in 2+1 dimensions spatial infinity is not simply
connected  so that, as
appreciated in [\HC], spinors that are asymptotically Killing may be
either  periodic or
anti-periodic at spatial infinity. However, knowledge of the behaviour
at  spatial infinity of an
asymptotically Killing spinor requires more than the knowledge of
whether it  is periodic or
antiperiodic; one must know its phase, which is specified by the
integer $n$.  We shall shortly
return to the significance of this fact.

A variant [\AD] of the ADM method can be used to find expressions as
surface  integrals of mass,
angular momentum and charge in asymptotically adS spacetimes of
arbitrary  dimension. In three
dimensions a simplification of this method is made possible by the
fact that  three-dimensional
adS (super)gravity can be viewed as a pure Chern-Simons theory
[\AT,\AW]. This  leads naturally to
a definition of mass and other `charges' in terms of holonomy, in
agreement  with other methods
[\BTZ,\MS]. We use this result and the methods of [\GH,\GKLTT] to
deduce that  the energy $E=M+1$
of an  asymptotically adS spacetime is bounded in terms of its angular
momentum and charge.
Specifically, for zero angular momentum and charge one finds that
$M\geq -n^2$, where the integer
$n$ specifies the type of asymptotically Killing spinor under
consideration.  The proof of this
bound depends on the asymptotically Killing spinor being non-singular
on a  Cauchy surface and satisfying a Dirac-like equation (the `Witten
condition'). It can happen that a particular
asymptotically adS spacetime will not admit such a spinor for certain
values  of $n$. An example is adS spacetime itself for which $|n|=1$
is  possible but $n=0$ is not; if it were we could prove
that $M\geq 0$ for this spacetime whereas in fact $M=-1$.  We shall
later  provide some evidence
for the conjecture that, for any asymptotically adS solution of the
Einstein  equations (with suitably well-behaved matter) for which
there  exists a non-singular Cauchy surface, it is
possible to find an asymptotically Killing spinor satisfying the
Witten  condition with {\it either} $n=0$ {\it or} $|n|=1$. If this is
true  then we need consider only $n=0$ and $|n|=1$,
since $|n|>1$ leads to a weaker bound. This would establish the
absolute  stability of adS spacetime, but the black hole vacuum might
be  unstable against semi-classical tunnelling. An
instability of this type has been shown to afflict the Kaluza-Klein
vacuum of  five-dimensional general relativity [\KK], for which the
positive  energy theorem fails for similar reasons.

This conjecture does not imply the non-existence of spacetimes with
$M<-1$; as  already mentioned, there is a supersymmetric spacetime
with  $M=-n^2$ for every integer $n$. However, the
supersymmetric spacetimes with $|n|>1$ have naked singularities that
prevent  one from finding a non-singular Cauchy surface and this may
be a  feature of all solutions of the field equations
with $M<-1$, for sufficiently well-behaved matter. We remark that
spacetimes  with naked singularities should not be considered
unphysical {\it  per se}. It is known that cosmic
censorship is false for asymptotically $(adS)_3$ spacetimes since
non-singular  matter can collapse to a naked singularity if $0>M>-1$
[\RM].  The singularity in this case is a conical one
with a positive deficit angle. We suggest that the appropriate
extension of  the cosmic censorship
hypothesis in this context is such as to permit naked singularities of
this  type but no others.

In certain respects, (2+1)-dimensional supergravity is similar to
(4+1)  dimensional supergravity for which it is known [\GKLTT] that
supersymmetric black holes can be viewed as limits of
self-gravitating solitons of supergravity coupled to matter. One of
the motivations for the work reported in this paper was to investigate
whether  a similar result holds for self-gravitating
solitons of the 2+1 dimensional (2,0) adS supergravity coupled to
matter. In  addressing this
question our first task is the construction of a suitable coupling of
matter  to (2,0) adS
supergravity. The problem of coupling scalar multiplets to
$N$-extended  Einstein supergravity
without a cosmological constant was solved in [\dWNT]. The inclusion
of a  cosmological term is
straightforward for $N=1$ (leading to a matter coupled (1,0) adS
supergravity  theory), and is
implicit in the $N=1$ superspace results of [\GGRS]. In contrast, the
inclusion of a cosmological
term for $N>1$, in the presence of matter, cannot be deduced simply
from the  results of [\dWNT]
because the field content  differs by the presence of the Chern-Simons
gauge  field(s). Taking the
$N=2$ flat space sigma-model as our starting point we construct a
locally  supersymmetric coupling
of this action to (2,0) adS supergravity. The target space of this
locally-supersymmetric sigma
model is K{\"a}hler. We choose it to be compact, in order to allow for
the  possibility of
sigma-model solitons, and we study the particular case of the Riemann
sphere  in detail. In the
absence of supergravity the soliton solutions are simply holomorphic
functions  on 2-space
considered as the complex plane (see, for example, [\SSS]).
Remarkably, this  continues to be
true, for an appropriate choice of complex coordinate, even when the
gravitational corrections are included, although the holomorphic
functions  must now satisfy a further restriction. For the
case of an $S^2$ target space this further restriction is that the
holomorphic function be homogeneous. These self-gravitating solitons
admit  Killing spinors and saturate an energy bound.
In this respect, self-gravitating solitons of three-dimensional
supergravity/matter theories are similar to those of five-dimensional
supergravity [\GKLTT], but the Einstein equation is now a
{\it non-linear} second-order ODE.  Although we have not been able to
solve  this equation
analytically, we argue that the solitons governed by this equation are
non-singular. We are also
able to determine their asymptotic behaviour; we find that the
solitons  always have a logarithmic
dependence on the natural radial coordinate, for large radius, but
such that  $E-2mJ$ is
well-defined, and bounded in terms of the charge $Q$, even though
neither $E$  nor $J$ is
separately well-defined. Thus, the asymptotic symmetry group of these
solutions is a proper subgroup of the adS group.

We conclude this introduction with a brief account of the pure (2,0)
adS  supergravity [\AT]. The field content consists of the metric
tensor,  $g_{\mu\nu}$, a complex Rarita-Schwinger field,
$\psi_\mu$, and an Abelian gauge field $A_\mu$. The action is
$$
S= \int\! d^3x\, [-{1\over2}eR +{i\over2}\varepsilon^{\mu\nu\rho}\bar\psi_\mu
{\cal D}_\nu \psi_\rho
 -2m\varepsilon^{\mu\nu\rho}A_\mu\partial_\nu
A_\rho + 4m^2 e]\ ,
\eqn\twoa
$$
where $m$ is a constant with units of mass which we may assume to be
positive.  The covariant
derivative ${\cal D}$ is defined by
$$
{\cal D}_\mu = D_\mu + 2im A_\mu -im\gamma_\mu
\eqn\twob
$$
where $D$ is the usual Lorentz-covariant derivative except that the
spin connection has the non-zero torsion
$$
T_{\mu\nu}{}^a \equiv D_{[\mu}e_{\nu]}{}^a = {1\over
4}\bar\psi_\mu\gamma^a \psi_\nu\ .
\eqn\twoc
$$
The action (1.1) is invariant under the supersymmetry transformations
$$
\eqalign{
\delta_\epsilon e_\mu{}^a &= {1\over4}\bar\epsilon\gamma^a\psi_\mu
+ c.c.\cr
\delta_\epsilon\psi_\mu &= {\cal D}_\mu\epsilon \cr
\delta_\epsilon A_\mu &= {1\over4} \left( \bar\epsilon \psi_\mu
+c.c.\right)\  .}
\eqn\twod
$$
We follow the conventions of [\AT], i.e. the metric signature is
`mostly  minus' and the gamma
matrices $\gamma^a$ are pure imaginary.

%%%%%%%%%%%%%%%%%%%%%%%%%%%%%%%Chapter2%%%%%%%%%%%%%%%%%%%%%%%%%%%%%%%%

\chapter{ Killing Spinors}

The condition for a purely bosonic field configuration of (2,0) adS
supergravity to preserve
at least one supersymmetry of (2,0) adS supergravity is that the equation
$$
\delta_\epsilon \psi_\mu =0
\eqn\threea
$$
admit a non-zero solution for the spinor parameter $\epsilon$.
Since this equation is linear in $\epsilon$ its consequences are
unchanged if $\epsilon$ is replaced by the {\it commuting} spinor $\kappa$.
Making this replacement, we have to solve
$$
{\cal D}_\mu\kappa =0
\eqn\threeb
$$
for non-zero $\kappa$.

An alternative form for the covariant derivative ${\cal D}$ is
$$
{\cal D}_\mu =\partial_\mu +iB_\mu{}^a\gamma_a + 2im A_\mu\ ,
\eqn\threec
$$
where
$$
B_\mu{}^a = {1\over4}\varepsilon^{abc}\omega_{\mu\; bc}-me_\mu{}^a
\eqn\threed
$$
is an $Sp(2; \R)\cong Sl(2;\R)$ gauge field. This is a reflection of
the fact  that $(adS)_3$
gravity can be viewed as a Chern-Simons theory for the adS group
$SO(2,2)\cong Sl(2;\R)\times Sl(2;\R)$ [1], and that for the extension to (p,0)
supergravity only one $Sl(2;\R)$ factor is supersymmetrized. Since the torsion
vanishes in a purely bosonic background, the $Sl(2;\R)$ gauge field $B$ can be
expressed entirely in terms of the dreibein as follows:
$$
B_\mu{}^a= {1\over4}\varepsilon^{abc}\big[ 2 e_b{}^\nu
\partial_{[\mu}e_{\nu]\, c} -
e_b{}^\nu e_c{}^\rho \partial_\nu e_\rho{}^d e_{\mu\, d}\big] - me_\mu{}^a\ .
\eqn\threee
$$
Thus, Killing spinors $\kappa$ are non-zero solutions of
$$
(\partial_\mu  +iB_\mu{}^a\gamma_a + 2imA_\mu )\kappa =0
\eqn\threef
$$
with $B_\mu{}^a$ as given above. This equation has the integrability condition
$$
(mG^\mu + S^\mu{}_a\gamma^a)\kappa =0
\eqn\threeg
$$
where
$$
\eqalign{
S^\mu{}_a &\equiv \varepsilon^{\mu\nu\rho}(\partial_\nu B_{\rho\, a} -
\varepsilon_{abc}B_\nu{}^b B_\rho{}^c)\cr
G^\mu &\equiv 2\varepsilon^{\mu\nu\rho}\partial_\nu A_\rho}
\eqn\threeh
$$
are the duals of the $Sl(2;\R)$ and $U(1)$ field strengths respectively.

In the absence of matter the Euler-Lagrange equations for the bosonic
fields  of (2,0) adS
supergravity are simply
$$
S^\mu{}_a =0 \qquad G^\mu =0\ ,
\eqn\threei
$$
so that \threeg\ is automatically satisfied by solutions of the source-free
equations. Since only the $(adS)_3$ spacetime with zero $U(1)$ charge
admits  the full
complement of Killing spinors, i.e. four for (2,0) supergravity (this
being  the real dimension of
the space of complex two-component spinors) we shall seek metric and
gauge  field configurations
admitting two Killing spinors\foot{This will of course include the
zero charge  anti-de Sitter
spacetime as a special case.}. We can implement this by requiring
$\kappa$ to  satisfy a condition
of the form $$ (1-\gamma^a b_a)\kappa=0 \eqn\threej
$$
for some complex functions $b_a$. Provided that
$$
b\cdot b=1
\eqn\threek
$$
this condition projects out half the components of $\kappa$.
This constraint on $\kappa$ can be solved by writing
$$
\kappa=N (1+\gamma^a b_a)\zeta_{{}_0}
\eqn\threel
$$
for arbitrary complex function $N$ and constant spinor $\zeta_{{}_0}$.
When \threel\ is substituted into the integrabilty condition
\threeg\ one finds that
$$
S^\mu{}_a + mG^\mu b_a + i\varepsilon_{abc}b^bS^{\mu\, b} =0\ .
\eqn\threem
$$
This is of course satisfied by any solution of the empty space
equations, but  we
wish to find {\it off-shell} configurations that admit Killing spinors. When we
subsequently consider the implications of the field equations our
Killing  spinor results will then be equally applicable whether we
consider  the field equations with or
without matter. Note that the integrability condition \threeg\ is
necessary  but {\it
insufficient} for the background to admit a Killing spinor, so after
analysing  the content of the
integrability conditions we must then return to the Killing spinor
equation  \threef, which will
restrict the function $N$ in \threel.

Rather than attempt to find the general solution of \threef\  we shall
find  solutions by making
the following ansatz for the frame one-forms
$e^a$:
$$
e^\uzero = f(r)dt \qquad e^\uone = h(r)dr \qquad e^\utwo = {u(r)\over r}dt +
rd\varphi\ ,
\eqn\threen
$$
where underlining indicates a frame index (i.e. $a=\uzero,\uone,\utwo$).
The metric is then given by
$$
ds^2 = \big[ f^2 - \left({u\over r}\right)^2\big] dt^2 -2udtd\varphi - h^2 dr^2
-r^2d\varphi^2\ .
\eqn\athreen
$$
This ansatz is suggested by the fact that the function $u(r)$ is
constant for  the black hole.
The $Sl(2,\R)$ one-form potential $B_a$ may now be computed from
\threee\ and the result is
$$
\eqalign{
B_\uzero &= \left( {u'\over 4rh} -mf\right)dt + {1\over 2h} d\varphi\cr
B_\uone &=\left[{1\over 4rf}\left({2u\over r} -u'\right) + mh\right] dr\cr
B_\utwo &= \left[ -{1\over 4h}\left( 2f' -{uu'\over r^2 f} +
{2u^2\over r^3  f}\right)
+ {mu\over r}\right]dt \cr
& \qquad +\left[ {1\over 4fh}\left( u' -{2u\over r}\right) + mr\right]
d\varphi\ .}
\eqn\threeo
$$
Our ansatz may now be shown to imply that certain
components of $S^\mu{}_a$ vanish. Specifically, one finds that
$$
S^r{}_\uzero = S^r{}_\utwo = S^t{}_\uone = S^\varphi{}_\uone =0\ .
\eqn\threep
$$
We now separate the integrability condition \threem\  into its various
components. Firstly we
have
$$
S^r{}_\uone=0\ ,
\eqn\threeq
$$
which, since $B_\mu{}^a$ depends only on $r$, is equivalent to
$$
B_{\varphi\utwo}B_{t\uzero} = B_{\varphi\uzero}B_{t\utwo}\ .
\eqn\threer
$$
Secondly, we have
$$
\qquad mG^\mu + S^\mu{}_a b^a =0\ ,
\eqn\threes
$$
and finally
$$
\eqalign{
(1-b_\uzero^2) S^{t\uzero} +(ib_\uone - b_\utwo b_\uzero) S^{t\utwo} &=0\cr
(1-b_\uzero^2) S^{\varphi\uzero} +(ib_\uone - b_\utwo b_\uzero)
S^{\varphi\utwo} &=0\ .}
\eqn\threet
$$

We now return to the Killing spinor equation. Given that $\kappa$ has
the form  of \threel, we
find by substitution into \threef\ that
$$
2m A_\mu -i\partial_\mu(\ln N) +b_aB_\mu^a =0
\eqn\threev
$$
and
$$
B_\mu{}^a - i\varepsilon^{abc} b_b B_{\mu\, c} - i\partial_\mu b^a
- b^ab_c B_\mu^c =0\ .
\eqn\threew
$$

Observe now that \threer\ implies
$$
\eqalign{
B_{t\utwo} &= \Lambda (r) B_{\varphi\utwo}\cr
B_{t\uzero} &= \Lambda (r) B_{\varphi\uzero} \ ,}
\eqn\threex
$$
for some function $\Lambda$ of $r$, which must be finite since
$B_{\varphi\uzero}\ne0$. It follows that the $t$ and $\varphi$
components of \threew\ are consistent only if
$$
(\partial_t - \Lambda\partial_\varphi)b_a =0 \ .
\eqn\threey
$$
Differentiating this equation with respect to $r$, and denoting this
derivative by a prime, we find, on the one hand, that
$$
(\partial_t - \Lambda\partial_\varphi)b_a' = \Lambda' \partial_\varphi b_a\ .
\eqn\threez
$$
On the other hand, the $r$ component of \threew\ yields
$$
\eqalign{
ib_\uzero' &= B_{r\uone}(ib_\utwo + b_\uzero b_\uone)\cr
ib_\uone' &= B_{r\uone}(1 + b_\uone^2)\cr
ib_\utwo' &= B_{r\uone}(ib_\uzero + b_\uone b_\utwo)\ ,}
\eqn\threeaa
$$
which we may use to eliminate $b_a'$ in \threeaa\ in terms of $b_a$
and  $B_{r\uone}$.
Since $B_{r\uone}$ depends only on $r$, and $b_a$ satisfies \threey, the
left hand side of \threez\ vanishes, and we conclude that
$$
\Lambda' \partial_\varphi b_a =0\ .
\eqn\threeab
$$
This leads to two fundamentally different ways of solving the Killing spinor
equation:
\vskip 0.3cm
\noindent{\bf Either}
\vskip 0.3cm
(i) $\Lambda$ is {\it constant}, in which case $b$ depends on $t$ and
$\phi$  only
through the combination $\hat \varphi \equiv \varphi +\Lambda t$, i.e.
$$
b_a = b_a(\hat \varphi, r)\ ;
\eqn\threeac
$$
{\bf or}
\vskip 0.3cm
(ii) $b_a$ is independent of both $t$ and $\varphi$, i.e.
$$
b_a = b_a(r)\ ,
\eqn\threead
$$
and $\Lambda(r)$ is unrestricted.
\vskip 0.3cm
Option (i) is the more difficult to analyse, partly because there is
in this  case a further
bifurcation of possibilities according to whether or not
$\partial_\varphi  b_\utwo$ vanishes. It
turns out that all supersymmetric spacetimes found via option (i) are
also  found via option (ii);
that this is possible is due to the fact that the function $N$ can be
$\varphi$-dependent. The converse is not true, however; option (ii)
leads to  some supersymmetric spacetimes that cannot be
found via option (i). We shall therefore omit the analysis of option
(i) and  proceed with option
(ii).

For either option, the $t$ and $\varphi$ components of \threew\ are
equivalent  so we need
consider only the latter. Since $\partial_\varphi b_a=0$ for option
(ii) these  reduce to the algebraic equations
$$
\eqalign{
B_{\varphi\uzero} (1-b_\uzero^2) +
B_{\varphi\utwo}(b_\uzero b_\utwo -ib_\uone) &=0\cr
B_{\varphi\uzero} (ib_\utwo - b_\uone b_\uzero) +
B_{\varphi\utwo}(b_\uone b_\utwo -ib_\uzero) &=0\cr
B_{\varphi\uzero} (ib_\uone +b_\uzero b_\utwo) -
B_{\varphi\utwo}(1+ b_\utwo^2) &=0 \ .}
\eqn\algebraic
$$
Solving the last of these equations for $b_\uone$ and substituting it
into the other two equations we find that
$$
\eqalign{
b_\uzero &= Rb_2 \pm \sqrt{1-R^2} \cr
b_\uone &= -i[ R \mp b_\utwo \sqrt{1-R^2}]\ ,}
\eqn\foura
$$
where
$$
R = {B_{\varphi \utwo}\over B_{\varphi\uzero}}
= {1\over 2f}\big( u' -{2u\over r}\big) + 2mhr \ .
\eqn\fourb
$$
Substituting this result into the equations \threeaa\ one discovers
that $R$  and
$b_\utwo$ satisfy the ordinary differential equations
$$
R' - 2(1-R^2) B_{r\uone} =0
\eqn\fourc
$$
and
$$
b_\utwo ' = \pm B_{r\uone}\sqrt{1-R^2} (1+b_\utwo^2)\ .
\eqn\fourd
$$
The equation for $R$ together with \threer\ is equivalent to the following two
equations that determine the functions $f$ and $h$ in terms of $u$ and
two  constants $\alpha$
and $\beta$:
$$
\eqalign{
4mhf &= 2\alpha - {u'\over r} \cr
f^2 &= \big(\alpha r - {u\over r}\big)^2 + \beta^2 \ .}
\eqn\foure
$$
The constant $\alpha$ must be real and cannot vanish if the spacetime is to be
asymptotic to anti-de Sitter space (or the black hole vacuum). By a
redefinition of the time coordinate $t$, we can effectively change
$\alpha$ to  any non-zero value, and the choice
$$
\alpha=2m
\eqn\ealpha
$$
is convenient. The constant $\beta^2$ must also be real. This would,
in  principle, allow $\beta$
to be pure imaginary but we shall seee shortly that $\beta$ must be real too.

One can now deduce the following useful formulae:
$$
\eqalign{
R = {\sqrt{f^2-\beta^2} \over f} \qquad &\qquad
\sqrt{1-R^2} = {\beta\over f} \cr
B_{r\uone} = {1\over 2f} \big(2m r -{u\over r}\big)' \qquad & \qquad
\Lambda \equiv {B_{t\uzero}\over B_{\varphi\uzero}} = {u'\over r} -2m \ .}
\eqn\fourf
$$
Using these formulae one can easily solve \fourd\ for $b_\utwo$ and
simplify  the
expressions for the other components of $b_a$. The result is
$$
\eqalign{
b_\uzero &={\sqrt{f^2-\beta^2}\over f}b_\utwo + {\beta\over f} \cr
b_\uone &= -i\Big({\sqrt{f^2-\beta^2}\over f} - {\beta\over f} b_\utwo\Big) \cr
b_\utwo &= {k_+ \sqrt{f+\beta} + k_-\sqrt{f-\beta}\over
k_- \sqrt{f+\beta} - k_+\sqrt{f-\beta}} }
\eqn\fourg
$$
where $k_\pm$ are arbitrary constants.

We now turn to the equation for the one-form $A$, \threev. If the
associated  abelian gauge
group is ${\R}$ then the gauge freedom of $A$ can be used to remove
the phase  of $N$. However, if
the gauge group is $U(1)$ then we must distinguish between `small' and
`large'  gauge transformations, i.e. between elements of $U(1)$ that
are  connected to the identity and those
which are not connected to the identity. Because the action for the
$U(1)$  gauge field $A$ is $m$
times the CS term, the charge $Q$ associated with $A$ is $$ Q= {m\over
\pi}  \oint_\infty A\ ,
\eqn\chargedef $$ as will be explained in more detail in section 4. A
large  gauge transformation will change the value of this charge, so
fixing  the charge restricts the gauge
freedom to small gauge transformations. In this case the phase of $N$
cannot necessarily be completely removed by a gauge transformation but
we can  use the freedom of small gauge
transformations to bring the function $N$ to the form
$$
N = a(r,\varphi) e^{{in\over 2}\varphi +
im\beta t}\ ,
\eqn\fourh
$$
where $n$ is an integer and $a(r,\varphi)$ is a real function. The t-dependent
term in the phase is chosen for later convenience. Taking real and
imaginary  parts of \threev\
now determines the one-form $A$ and the function $a(r,\varphi)$. It is
at this  point that one
discovers that $\beta$ must be purely real rather than purely
imaginary  because $a(r,\varphi)$ would otherwise be a non-periodic
function  of $\varphi$. For real $\beta$ one finds that
$$
A= - \Big({\beta u'\over 2(4m r -u')}\Big)dt -
\Big({\beta r\over 4m r - u'} +{n\over 4m}\Big)d\varphi
\eqn\fouri
$$
and
$$
a = a(r) \equiv k_- \sqrt{f+\beta} - k_+ \sqrt{f-\beta}\ .
\eqn\fourj
$$

Combining \threel\ with \fourg\ now yields the following
result for the Killing spinor $\kappa$:
$$
\eqalign{
\kappa &= e^{{in\over 2}\varphi + im\beta t}[k_- \sqrt{f+\beta} -
k_+ \sqrt{f-\beta}] \cr
\times &\Bigg\{\Big[ 1 + {1\over f} \big( \beta\gamma^\uzero -
i\sqrt{f^2-\beta^2}\gamma^\uone\big)\Big] +
b_\utwo\gamma^\utwo \Big[ 1 - {1\over f} \big( \beta\gamma^\uzero -
i\sqrt{f^2-\beta^2}\gamma^\uone\big)\Big]\Bigg\}\zeta_0\ . }
\eqn\fourk
$$
It is convenient to normalize the constant spinor $\zeta_0$
such that $$ \zeta_0^\dagger \zeta_0 =1\ .
\eqn\normzeta
$$
In addition, we may require without loss of generality that $\zeta_0$
satisfy  $P\zeta_0=\zeta_0$
for some constant projection matrix, $P$, that projects out two
components of  $\zeta_0$, since
(for fixed $k_\pm$) the real dimension of the space of Killing spinors
$\kappa$ is two, not four,
despite the fact that $\zeta_0$ has two complex, and hence four real,
components. A convenient choice is
$$
\Big[k_+^2 + k_-^2\Big]^{-1}\Big[ (k_-^2 -k_+^2)\gamma^\uzero + 2ik_+k_-\,
\gamma^\uone\Big]\zeta_0 = \zeta_0\ .
\eqn\anormzeta
$$
Given that $\zeta_0$ obeys these restrictions, we find that
$$
\eqalign{
\bar\kappa\kappa &= 8ik^2\beta  \cr
\bar\kappa \gamma_\uzero\kappa &= 8ik^2 f  \cr
\bar\kappa \gamma_\uone\kappa &= 0\cr
\bar\kappa\gamma_\utwo\kappa &= 8i k^2 \sqrt{f^2-\beta^2}\ ,}
\eqn\aeightk
$$
where $k^2 = (k_+^2 + k_-^2)$; these results will prove useful later.

The metric admitting the Killing spinors \fourk\ is
$$
ds^2 = \big((2mr)^2 -4m u + \beta^2\big)dt^2 -2u dtd\varphi
-{\big[1 -\left({u'\over 4mr}\right)\big]^2\over
\big[\big(2m r - \left({u\over r}\right)\big)^2 + \beta^2\big]} dr^2
-r^2d\varphi^2\ .
\eqn\fourl
$$
Notice that the requirement that there exist Killing spinors leaves
undetermined the function $u$ and the constant $\beta$. These are to
be  determined by the field equations and the boundary conditions.

%%%%%%%%%%%%%%%%%%%%%%%%%%%%%%%Chapter3%%%%%%%%%%%%%%%%%%%%%%%%%%%%%%%%

\chapter{ Supersymmetric vacuum spacetimes}

In section 6 we shall consider the field equations with a sigma model
soliton  source. Here we shall consider the simpler source-free case.
Inserting the metric \fourl\ into the source-free
Einstein equation (with cosmological constant) we find that $$
\Big({u'\over r}\Big) = {\rm const.}
\eqn\fourm
$$
By a redefinition $\varphi \rightarrow \varphi + {\rm const}.\times
t$, we can  arrange for the
integration constant on the right hand side of \fourm\ to vanish, in
which  case
$$
u= {1\over2}J
\eqn\fourn
$$
for {\it constant} $J$. The metric, with $\alpha=2m$, is then
$$
ds^2 = [(2mr)^2 -2mJ +\beta^2) dt^2 - J dt d\varphi - {4r^2 dr^2\over (4m r^2
-J)^2 + 4\beta^2 r^2} - r^2d\varphi^2\ .
\eqn\fouro
$$
By comparison with the black hole solution of [\BTZ] we see that $J$
is the  total angular
momentum of this spacetime, and that the total mass, relative to the
black  hole vacuum, is $M=
2mJ -\beta^2$. We shall confirm these identifications in the following
section. The gauge-field one-form is
$$
A=  -{1\over 4m}(\beta + n)d\varphi\ ,
\eqn\fourp
$$
and from \chargedef\ we see that the associated charge is
$$
Q= -{1\over2}(\beta +n)\ .
\eqn\charge
$$
Using this relation between $\beta$ and $Q$, the total mass may be
expressed  in terms of $J$ and
$Q$ as
$$
M= 2mJ -(2Q+n)^2\ .
\eqn\massform
$$

Although $\beta$ need not be an integer, one must choose $\beta=-n$ to
get  $A=0$ and so recover
the spacetimes considered previously in the context of the purely
Einstein theory. In this case the charge $Q$ vanishes and the metric is
$$
ds^2 = [(2mr)^2 -2mJ +n^2] dt^2 -J dt d\varphi - {4r^2 dr^2\over (4m r^2
-J)^2 + 4n^2 r^2} - r^2d\varphi^2\ .
\eqn\metrica
$$
When $n=0$ this metric represents a spacetime of mass $M=2mJ$. For
{\it  positive} $J$ (and $n=0$) the mass $M$ is positive and the
metric is  that of the extreme black hole, shown previously to admit
Killing  spinors [\HC]. A version of this supersymmetric extreme black
hole  metric with positive mass exists for either sign of $J$ because
the  transformation $t\rightarrow -t$ in \metrica\ effectively changes
the  sign of $J$ without changing the mass $M$, so the extreme black
hole  metric may be more generally written as [\BTZ]
$$
ds^2 = \big[(2mr)^2 -2m|J|\big] dt^2 -J dt d\varphi - {4r^2 dr^2 \over
(4m r^2  -|J|)^2 }
- r^2d\varphi^2\ .
\eqn\ametrica
$$
Note that the metric \metrica\ continues to admit Killing spinors when
$J$ is  negative, so the metric
$$
ds^2 = \big[(2mr)^2 +2m|J|\big] dt^2 -J dt d\varphi
- {4r^2 dr^2 \over (4m r^2 +|J|)^2} - r^2d\varphi^2\ ,
\eqn\bmetrica
$$
with {\it negative} mass $M=-2m|J|$ is {\it also} supersymmetric!
Unlike the  extreme black hole,
however, it has a {\it naked} cusp-like singularity at $r=0$. For
either sign  of the mass, the
Killing spinor $\kappa$ is given by
$$
\kappa = \sqrt{2mr - {M\over 4mr}}\ \big(1-i\gamma^\uone \big)\psi_0
\eqn\spinbh
$$
where $\psi_0 =\big[ (k_- -k_+) +(k_- + k_+)\gamma^\utwo\big]\zeta_0$
is a  constant spinor.
In these cases, the freedom represented by $k_\pm$ is clearly not
independent  of the freedom to
choose $\psi_0$ and the Killing spinors span a space of real dimension
2. When  $M>0$ the Killing
spinor $\kappa$ vanishes at $r=\sqrt{M/8m^2}$, which is the horizon of
the  black hole. For either sign of $M$ the Killing spinor is singular
at  $r=0$, but so also is the metric.

Consider now the static spacetimes with $J=0$. The metric is
$$
ds^2 = [(2mr)^2 +n^2] dt^2 - {1\over \Big[(2mr)^2 + n^2\Big]}dr^2 -
r^2d \varphi^2\ .
\eqn\cmetrica
$$
For $n=0$ this is the metric of the black hole vacuum:
$$
ds^2 = r^2\Big[ d(2mt)^2 - d\varphi^2\Big] - {dr^2\over r^2}
\eqn\bhvac
$$
which would be one form of the adS metric, which has a non-singular
Killing  horizon at $r=0$,
were it not for the fact that $\varphi$ is an angular variable. The
identification of $\varphi$
with $\varphi + 2\pi$ creates a conical singularity at the horizon, so
the  black hole vacuum has
a null singularity at $r=0$.

Observe that a necessary condition for a Killing spinor $\kappa$ to be
non-singular at $r=0$ is that
$$
\kappa \sim r^{\sigma\over2} e^{\pm {in\over 2}\varphi}\kappa_0 \qquad
(t=0,\,  r\rightarrow 0)
\eqn\spinorigin
$$
for some constant (non-zero) spinor $\kappa_0$ and non-negative
integer  $\sigma$. Moreover, if
$\sigma=0$ the spinor $\kappa$ will still be singular at $r=0$ unless $|n|=1$.

When $|n|=1$, the metric \cmetrica\ is that of adS space. The
corresponding  Killing spinors
have $\sigma=0$ and are therefore non-singular at $r=0$, as required
since the  singularity
of the metric at $r=0$ is merely a coordinate singularity in this case.
When $|n|>1$, the Killing spinors $\kappa$ still have the property
\spinorigin\ with $\sigma=0$. They are therefore singular at $r=0$,
but this  is not problematic because the metric
is also singular there. To see this we observe that when $|n|>1$ we
may  introduce the new variables
$$
\hat r = {r\over n}\ ,\qquad  \hat\varphi = n\varphi\ ,
\eqn\newvar
$$
in terms of which the metric \cmetrica\ is again adS, but the angular
variable $\hat\varphi$ is now identified with period $2|n|\pi$ instead
of  $2\pi$, which implies a conical singularity at $r=0$ with negative
deficit  angle $\delta=-2(|n|-1)\pi$. The `point'
$r=0$ must therefore be excluded from the spacetime. It seems that
such  spacetimes are unphysical.

When $Q\ne0$ most of the above discussion still applies if $Q$ is
half-integral, but with $n$ replaced by $n+2Q$. Note, however, that
although  the metric is adS for $J=0$ and $|n+2Q|=1$, and
hence non-singular, the gauge field $A$ is singular at the origin if
$Q\ne0$,  so that one cannot have a completely non-singular `charged'
adS  spacetime. In contrast, when $J=0$ and $|n+2Q|=0$,
so that the metric is that of the black hole vacuum, it is still
possible to  have $Q=-{n\over2}$ for $n\ne0$ without incurring a
singularity  of the gauge field. This is due to the fact that the
spatial sections of the black hole vacuum are not simply connected.

There is no necessity for $Q$ to be half-integral, however. Consider,
for  example, the static supersymmetric spacetimes with $J=0$ and
$n=0$, for  which $Q= -{1\over2}\beta$. Because of the
freedom to adjust $n$ we may assume that
$$
0\le \beta \le 1\ .
\eqn\rangebeta
$$
The metric is
$$
ds^2 = [(2mr)^2 +\beta^2) dt^2 - {dr^2 \over \Big[(2m r)^2 +
\beta^2\Big]} -  r^2d\varphi^2\ ,
\eqn\charged
$$
and the mass is $M=-\beta^2=-4Q^2$. These spacetimes have been
considered in  [\DJ]; they have a conical singularity at the origin
with  deficit angle $\delta= 2\pi(1-\beta)= 2\pi (1+
\sqrt{M})$. They are nevertheless physical because they may form from
an  initially non-singular shell of matter that collapses to the
origin [\RM].  They may be regarded as point particle
solutions and in this sense are not truly vacuum solutions. Here we
have shown  that these solutions are supersymmetric, for appropriate
$U(1)$  charge, in the context of (2,0) adS supergravity.

%%%%%%%%%%%%%%%%%%%%%%%%%%%%%%%Chapter4%%%%%%%%%%%%%%%%%%%%%%%%%%%%%%%%

\chapter{ An energy bound for (2+1) adS gravity}

The total mass and angular momentum of a (3+1)-dimensional asymptotically adS
spacetime were defined by Abbott and Deser [\AD] using a modification
of the  ADM procedure for asymptotically-flat spacetimes. Although the
same analysis can be directly taken over for (2+1)-dimensional
asymptotically-flat spacetimes, a similar but much
simpler analysis is available that exploits the Chern-Simons
formulation of  adS gravity in 2+1 dimensions.

Consider first an arbitrary semi-simple Lie algebra. Let $K_\mu$ be the Lie
algebra-valued CS gauge field with action
$$
S= \int d^3x \tr \Big[ \varepsilon^{\mu\nu\rho}( K_\mu\partial _\nu
K_\rho +  {2\over3}
K_\mu K_\nu K_\rho) + K_\mu j^\mu (\phi) \Big]\ ,
\eqn\afivea
$$
where $j^\mu(\phi)$ is a Lie algebra valued current for the `matter'
fields,  $\phi$. The field
equation of $K_\mu$ is
$$
\varepsilon^{\mu\nu\rho}K_{\mu\nu} = -j^\mu (\phi)
\eqn\fivea
$$
where $K_{\mu\nu}=\partial_\mu K_\nu -\partial_\nu K_\mu$ is the
field-strength tensor. We now
split $K_\mu$ into a background $\bar K_\mu$ and a (not necessarily
small)  perturbation $\Delta K_\mu$, i.e.
$$
K_\mu= \bar K_\mu + \Delta K_\mu  \ .
\eqn\fiveb
$$
The background is taken to satisfy the source-free equations, i.e.
$$
\bar K_{\mu\nu}=0\ .
\eqn\fivec
$$
We can now rewrite the field equation \fivea\ as
$$
\varepsilon^{\mu\nu\rho} \bar D_\nu (\Delta K)_\rho =
-j^\mu_{tot.}(\phi,  \Delta K)
\eqn\fived
$$
where $\bar D$ is the covariant derivative constructed from the
background  field. Thus, the left hand side of \fived\ is {\it linear}
in  $\Delta K$; the non-linear terms have been moved to the
right hand side and included in the `total' current $j_{tot.}$. If we
now  suppose that the background $\bar K$ is such as to admit a
Lie-algebra  valued scalar $\xi$ that is background
covariantly constant, i.e.
$$
\bar D_\mu \xi =0\ ,
\eqn\fivee
$$
then the Lie algebra invariant vector density
$$
J^\mu = {1\over 2\pi}\tr (j_{tot.}^\mu \xi)
\eqn\fivef
$$
is conserved, i.e. $\partial_\mu J^\mu =0$. The field equations in the
form  \fived\ can now be
used to re-write this as the identically conserved current
$$
J^\mu = -{1\over 2\pi} \varepsilon^{\mu\nu\rho}\partial_\nu \tr( \Delta K_\rho
\xi)\ .
\eqn\fiveg
$$
The associated charge is the surface term
$$
Q(\xi) = -{1\over 2\pi} \oint \tr (\Delta K\xi)\ ,
\eqn\fiveh
$$
where $\Delta K$ is the one-form with components $\Delta K_\mu$ and
the  integral is taken over a `circle at infinity'. We can now define
$Q(\xi)$  to be the total charge on a spacelike hypersurface
associated with a  Lie algebra valued scalar that approaches $\xi$
asymptotically. Clearly, $K$ must also approach $\bar K$
asymptotically. Thus,  the ADM procedure applied to
(2+1)-dimensional CS gauge theories leads to an identification of
charge as  the holonomy of an
asymptotic $U(1)$ connection.

To apply this result to the $U(1)$ gauge field $A$ of the (2,0) adS
supergravity we should take into account that the normalization of the
CS  action for this field differs from the canonical
normalization by a factor of $2m$. Thus, the associated charge $Q$ is
$$
Q= {m\over \pi} \oint \, A\ ,
\eqn\afiveh
$$
as claimed in the previous section. To obtain similar formulae for the
energy  and angular momentum of asymptotically $(adS)_3$ spacetimes,
we recall  that the equations of adS gravity are
equivalent to CS equations for the $Sl(2;\R)\times Sl(2;\R)$ one-forms,
$$
B_{(\pm)}^a = {1\over4}\varepsilon^{abc}\omega_{ab} \pm m e^a\ ,
\eqn\afivei
$$
associated with the generators $J^a \pm (2m)^{-1}P^a$, where $J^a=
{1\over2}\varepsilon^{abc}J_{bc}$ [\AT]. The normalization of the
charges in  \fiveh\ assumes a canonical normalization for the CS
action. The  field equations of adS gravity are obtainable as
the Euler-Lagrange equations of any linear combination of the CS terms
for the  two $Sl(2;\R)$ factors of the adS group, but the standard
Einstein-Hilbert form of the action is found by taking
$(2m)^{-1}$ times the {\it difference} of the CS terms for $B_{(+)}^a$
and  $B_{(-)}^a$ [\AT,\AW]. This means that we should identify the
charges  constructed from $B_{(\pm)}^a$ and an
appropriately normalized background-covariantly constant
future-directed  timelike vector $\xi^a$ with $\pm 2m[J \pm
(2m)^{-1}E]= E\pm  2mJ$, where $E$ is the energy and $J$ is the angular
momentum. Thus,
$$ E \pm 2m J = -{1\over 2\pi}\oint\! \Delta
B^a_{(\pm)} \xi_a  \eqn\fivej
$$
where $$ \Delta B^a_{(\pm)} = (B^a_{(\pm)} - \bar B^a_{(\pm)}) \eqn\fivek
$$
and the adS background gauge potentials $\bar B^a_{(\pm)}$ are given by
$$
\eqalign{
\bar B^{\uzero}_{(\pm)} &= -m\sqrt{(2mr)^2 +1}(dt \pm {1\over 2m}d\varphi)\cr
&= -2m^2 r\Big[1+ {1\over 8m^2r^2} + O({1\over r^4})\Big](dt \pm {1\over
2m}d\varphi)\cr
\bar B^\uone_{(\pm)} &= - {m\over \sqrt{(2mr)^2 + 1}} dr = -{1\over 2r}\Big[
1+O({1\over r^2})\Big] dr\cr
\bar B^{\utwo}_{(\pm)} &= -2m^2 r (dt \pm {1\over 2m}d\varphi) \ .}
\eqn\fivel
$$
The background covariantly-constant vector $\xi$ may be
expressed in terms of an adS Killing spinor $\kappa$ through the relation
$$
\xi^a = -{i\over2} \bar\kappa \gamma^a \kappa\ .
\eqn\afivel
$$
Note the definition
$$
\bar\kappa \equiv i\kappa^\dagger \gamma^\uzero\ ,
$$
and the identity
$$
(\bar\kappa\gamma^a\kappa)(\bar\kappa\gamma_a\kappa) \equiv
(\bar\kappa  \kappa)^2\ ,
\eqn\fivem
$$
which ensure that $\xi$ is future-directed timelike provided
$\bar\kappa \kappa$ is non-zero, a
condition that is satisfied for the adS background.  Using the results
of  \aeightk\ with
$\beta=1$, we find that $$
\eqalign{
\xi_0 &= 4k^2\sqrt{(2mr)^2+1} = 8mr\, k^2  + O({1\over r})\cr
\xi_1 &= 0\cr
\xi_2 &= 8mr\, k^2 \ .}
\eqn\bfivel
$$
To fix the constant $k$ we shall evaluate the right hand side of
\fivej\ for  the general black
hole solution of [\BTZ]. In our notation the black hole metric is
$$
\Big[ (2mr)^2 -M\Big] dt^2 - J d \varphi dt - {dr^2\over \Big[ (2mr)^2 -M +
{1\over 4r^2}J^2\Big]} - r^2 d\varphi^2
\eqn\BHmetric
$$
and, as pointed out in [\BTZ], the special case of $J=0$ and $M=-1$ is anti-de
Sitter spacetime, so we expect that $E=M+1$. Except in the extreme,
$M= 2mJ$,  case, this metric is not included in our ansatz of \threen\
because  the factor multiplying $dr^2$ in \BHmetric\ is
not positive definite for all $r$. However, this factor is positive
for large  $r$ so we may use our previous results to determine the
asymptotic  behaviour of $\Delta B^a_\varphi$.  We find that
$$ \eqalign{ \Delta B^\uzero_\varphi &= -{M+1\over 8mr} + O({1\over r^2})\cr
\Delta B^\utwo_\varphi &= {J\over 4r} + O({1\over r^2})\ . }
\eqn\asymb
$$
Combining this with \bfivel\ we verify the formula \fivej\ with $k=1$ and, as
expected, $E=M+1$. Let us note here for future reference that, for
$t=0$, the  asymptotic form of the Killing spinor \fourk\ as $r\rightarrow
\infty$ is
$$
\kappa \sim \sqrt{2mr}\, e^{i{n\over2}\varphi}\kappa_\infty \ ,
\eqn\asymkappa
$$
where
$$
\kappa_\infty = (1-i\gamma^\uone)\big[(k_- -k_+) + (k_- +
k_+)\gamma^\utwo\big]\zeta_0\ .
\eqn\kapinf
$$

We shall now use the above results to establish a lower bound on the
adS  energy $E$ following the Witten-Nester proof [\Witone] of the
positivity  of the ADM energy of asymptotically flat 3+1
dimensional spacetimes and its generalization to asymptotically adS
3+1  dimensional spacetimes [\GHW]. We shall consider asymptotically
adS  spacetimes admitting a spinor $\chi$ that is
asymptotic to one of the Killing spinors found previously. It follows
that,  asymptotically as $r\rightarrow \infty$,
$$
\eqalign{
-{i\over2}\bar\chi\gamma^a\chi  &\sim \xi^a \cr -{i\over8}
\bar\chi \chi &\sim 1 \ .} \eqn\asymchi
$$
The precise formulation of the asymptotic condition will be left until
later.  To avoid the proliferation of $\pm$ signs in the formulae to
follow we  shall assume that $J$ is non-negative.
In this case it will be sufficient to consider the `modified'
Nester-type  tensor
$$
\hat E^{\mu\nu}= -{i\over 4}\bar\chi\gamma^{\mu\nu\rho}
{\cal D}_\rho\chi \ + c.c.
\eqn\fiver
$$
where ${\cal D}$ is the covariant derivative of \threec.
Using \fivel\ and \asymchi, it is easily seen that \fivej\ can be rewritten as
$$
E - 2mJ =  {1\over 4\pi} \oint dS_{\mu\nu} \hat E^{\mu\nu} -
\Big[{1\over 8\pi} \oint \bar\chi \bar D\chi + c.c.\Big] -4Q \ ,
\eqn\fiveo
$$
where $\bar D$ is the adS background covariant derivative, and
$dS_{\mu\nu}$ is the dual of the line element of the circle at infinity.

Assuming now that the circle at infinity is the only boundary of a
spacelike two-surface with dual surface element $dS_\mu$, we have from
Gauss'  law that
$$
\eqalign{
{1\over4\pi}\oint\! dS_{\mu\nu} \hat E^{\mu\nu} &= {1\over 2\pi}\int\!
dS_\mu\, D_\nu\hat E^{\mu\nu}\cr
&= -{i\over4\pi}\int\! dS_\mu \big\{ \overline{{\cal
D}_\nu\chi}\gamma^{\mu\nu\rho}{\cal D}_\rho\chi - e^{-1}\bar\chi
(mG^\mu + S^\mu{}_a\gamma^a)\chi \big\} \ . }
\eqn\fivet
$$
At this point we need the field equations for the metric and the
abelian gauge  field. In order
to allow for sources we take these equations to be
$$
S^\mu{}_a = -{1\over2}T^{\mu\nu} e_{a\nu} \qquad \qquad G^\mu =
{1\over  2m}J^\mu\ ,
\eqn\sixa
$$
where $T^{\mu\nu}$ and $J^\mu$ are the matter stress tensor and $U(1)$ current
respectively. Then \fivet\ becomes
$$
{1\over4\pi}\oint\! dS_{\mu\nu} \hat E^{\mu\nu}
= -{i\over 4\pi}\int\!dS_\mu\big\{  \overline{{\cal
D}_\nu\chi}\gamma^{\mu\nu\rho}{\cal D}_\rho\chi +{1\over2}e^{-1}\bar\chi
\big(T^{\mu a}\gamma_a - J^\mu \big)\chi \big\}\ .
\eqn\sixb
$$
The second term in the integral on the RHS of this equation is
non-negative  provided that the vector field with components
$-i\bar\chi\Big(T^\mu{}_a\gamma^a -J^\mu\Big)\chi$ is non-spacelike
and future directed for all spinors $\chi$. This condition is
trivially  satisfied in the absence of matter and it is also satisfied
by the  currents of the supersymmetric sigma-model, as we shall
show later. The first term on the RHS of \sixb\ can be shown to be
positive  in the standard way provided that the spinor $\chi$
satisfies the  Witten-type condition $$ \dtwoslash\chi \equiv
{}^{(2)}g^{ij} \gamma_i {\cal D}_j \chi =0\ , \eqn\Witten $$ where
${}^{(2)}g^{ij}$ is the inverse of the spatial 2-metric.

It can be shown that, at least {\it locally}, there exists a solution
of  \Witten\ with the asymptotic form
$$
\chi \sim \sqrt{2mr} e^{{\mp i\nu\over 2}\varphi}\kappa_\infty \mp
\gamma^{\utwo}\Big[
{\nu\over2} + Q\Big]{ e^{{\mp i\nu\over2}\varphi}\over
\sqrt{2mr}}\kappa_\infty
\eqn\neqone
$$
for any integer $\nu$. Note that this spinor has the required
asymptotic  property \asymchi. To prove the existence of this solution
one  writes $\chi =\chi_\infty + \chi_1$ where $\chi_\infty$ equals
the right  hand side of \neqone\ and $\chi_1$ is assumed to fall off faster
than $1/\sqrt {r}$ as $r\rightarrow \infty$. One can then prove by a
variant  of the original argument of [\Witone] that the operator
$\dtwoslash$  has no zero eigenvalues on the space of
functions with the asymptotic fall off of $\chi_1$. It then follows
that there  exists, at least
locally, a solution of the form $\chi=\chi_\infty + \chi_1$ that is
determined  by $\chi_\infty$. The form of $\chi_\infty$ can be
determined by  solving the Witten condition for large $r$ and the
result is that given in \neqone. There may be global obstructions that
nevertheless prevent the existence of the required solution of the
Witten  condition. We shall return to this question
below. Given the absence of global obstructions, we have now shown
that the  first term in the integral on the RHS of \sixb\ is also
non-negative  and hence that the LHS is non-negative. Using
this information in \fiveo, we conclude that $$ E - 2mJ \geq
\Big[{1\over  8\pi} \oint \bar\chi\bar D\chi + c.c.\Big] -4Q \ ,
\eqn\fiveo $$  which can be saturated only if ${\cal D}\chi =0$,
i.e. only if $\chi$ is a Killing spinor. The line integral of
$\bar\chi\bar  D\chi$ may be computed using \neqone. It vanishes only
when  $\nu=1$ and $Q=0$ and the final result is such that
\fiveo\ is equivalent to
$$
E - 2mJ \geq 1 - \big( 2Q + \nu\big)^2
\eqn\neqthree
$$
By an appropriate change of sign in the covariant derivative in $\hat
E^{\mu\nu}$ one can also prove this inequality for the opposite sign
of $J$;  we thus derive the bound
$$
M \geq |2mJ| - \big( 2Q + \nu\big)^2\ .
\eqn\neqfour
$$
This is our main result concerning energy bounds in three-dimensional adS
spacetime. Note that for $\nu=n$ the bound is saturated by the
spacetimes  \fouro\ labelled by the integer $n$ in section 2. As we
saw there,  these solutions indeed admit Killing spinors.

Consider the special case for which $Q=0$. In this case the strongest
bound on  the energy is found by choosing $\nu=0$, in which case we have that
$$
M\geq 2m|J|\ ,
\eqn\neqfoura
$$
which was also found in [\MS] by other methods.
For $J=0$ this is saturated by the black hole vacuum which admits a
Killing  spinor having the asymptotic form of \neqone\ with $\nu=0$. However,
there may be global obstructions to the existence of a solution to the
Witten  condition that invalidate the bound. In fact, this is
necessarily the  case since the bound \neqfoura, for $J=0$,
is violated by adS space, which is non-singular and has $M=-1$ [\BTZ].
Clearly, the assumption of the absence of a global obstruction to the
existence of a solution to the Witten condition with
the asymptotic behaviour of \neqone\ must be false for adS space if
one  assumes that $\nu=0$. This accords with the fact that the Killing
spinor  of adS space has the asymptotic form \neqone\
but with $\nu=1$. For spacetimes for which $\nu=1$ gives the correct
asymptotic behaviour of globally-defined spinors one instead finds the
bound  $M\geq -1$, which is saturated by the adS
spacetime.

An instructive (although non-supersymmetric) example is provided by
the  interior metric of a homogeneous circularly symmetric collapsing
dust  ball:
$$
ds^2= dt^2 - a(t)^2\Big[ {dr^2\over (1-r^2)} + r^2d\theta^2\Big]\ .
\eqn\dustball
$$
The frame one-forms can be chosen to be
$$
e^\uzero = dt \qquad e^\uone = {a\, dr\over \sqrt{1-r^2}} \qquad
e^\utwo = ar  d\varphi\ ,
\eqn\frames
$$
from which we compute that
$$
\eqalign{
B_\uzero &= -mdt +{1\over2}\sqrt{1-r^2} d\varphi \cr
B_\uone &= am\sqrt{1-r^2}dr \cr
B_\utwo &= amrd\varphi \ .}
\eqn\bees
$$
The metric is non-singular at $r=0$ and Cauchy surfaces include this
point. It  follows that the spinor $\chi$ satisfying the Witten
condition must  be non-singular at $r=0$ and, as stated
previously, this restricts $\chi$ to have the behaviour
$$
\chi \sim r^{\sigma\over2} e^{i{\nu\over2}\varphi} \chi_0\qquad
(t=0,\,  r\rightarrow 0)
\eqn\chizero
$$
for integer $\nu$, non-negative integer $\sigma$ and non-zero constant
spinor  $\chi_0\,$.
Demanding that $\chi$ of this form satisfy \Witten\ one finds that
$$
(\sigma +1)\chi_0 = -\nu\gamma^\uzero\chi_0\ ,
\eqn\achizero
$$
which implies that
$$
(\sigma + 1)^2 = \nu^2\ .
\eqn\achi
$$
We see immediately that $\nu=0$ is not possible since $\sigma$ cannot
be  negative. The lowest admissable value of $|\nu|$ is $|\nu|=1$,
which  corresponds to $\sigma=0$. Linearity of \Witten\
and continuity imply that the phase of $\chi$ as $r\rightarrow \infty$
remains  equal $\nu\phi$, so the integer $\nu$ in \chizero\ is the
same as the  integer $\nu$ in \neqone\ which, as we have
now shown, cannot vanish. This is a satisfactory conclusion because
had a  non-singular spinor $\chi$ been allowed for $\nu=0$ we could
have used  it to derive the classical bound $M\geq 0$,
which we know to be false. As things stand, we may choose $|\nu|=1$
and this  allows the derivation of the classical bound $M\geq -1$. We
consider  this example as evidence for the conjecture that global
obstructions  to solutions of the Witten condition \Witten\ never exclude
(under the conditions discussed earlier) {\it both} $\nu=0$ and
$|\nu|=1$. If  this is true then the validity of the bound $M\geq -1$
extends  to the quantum theory. This would establish the
absolute stability of the adS vacuum. The status of the black hole
vacuum is  quite different since it saturates the weaker $\nu=0$ bound
$M\geq  0$. It seems likely that the black hole
vacuum will be unstable to decay by quantum tunelling for rather
similar  reasons to those that
lead to an instability of the 4+1 dimensional KK vacuum [\KK].

%%%%%%%%%%%%%%%%%%%%%%%%%%%%%%%Chapter5%%%%%%%%%%%%%%%%%%%%%%%%%%%%%%%%

\chapter{ (2,0) adS Supergravity sigma models}

The fields of the $N=2$ sigma-model action consist of the scalar fields
$\phi^I$, which are maps from spacetime to the $2n$-dimensional target
space  $\cM$, and the complex sigma-model fermions $\lambda^i$,
($i=1,\dots,n$). N=2 supersymmetry requires the target space to be
K{\" a}hler. That is, there must exist a closed two form $\Omega$
whose  components are related to a complex structure $\Omega^I{}_J$ on
$\cM$ by
$$
\Omega_{IJ} \equiv g_{IK}\Omega^K{}_J\qquad I,J,K =1,2,\dots,{\rm dim}\cM\ .
\eqn\sevena
$$
We note here that $d\Omega=0$ implies that $\Omega_{IJ}$ can be expressed {\it
locally} in terms of a vector $X$ as
$$
\Omega_{IJ} = X_{J,I} -X_{I,J} \ .
\eqn\sevenb
$$
The vector $X$ cannot be globally-defined on a compact K{\" a}hler
manifold  because in that case $\Omega$ is not exact. Since the
holonomy of  the (torsion free) affine connection of a
$2n$-dimensional K{\" a}hler manifold is $U(n)$ it is possible to
introduce a  set of complex frame one-forms $\{f^i,\, i=1\dots,n\}$ in
the  fundamental representation of $U(n)$, with components $f_I{}^i$.
Their  complex conjugates $\{ f_i\}$, with components $f_{I\, i}$ transform
in the $\bar n$ representation of $U(n)$. The metric $g$ and two-form
$\Omega$  can now be expressed in terms of the components of the frame
one-forms, i.e. the vielbein, as follows:
$$
g_{IJ} - i\Omega_{IJ} =
2f_I{}^i f_{J\, i}  \ .
\eqn\sevenc
$$
The introduction of the vielbein is convenient because it allows the
supersymmetry transformations involving the sigma-model fields to be
expressed  in terms of a single complex spinor parameter. Otherwise,
the  second-supersymmetry transformation, unlike the first, is
expressed in  terms of the complex structure and this obscures the fact
that both are really on the same footing.

The inverse complex vielbein $f^I{}_i$, with complex conjugate
$f^{I\, i}$, is defined by
$$
\eqalign{ f^I{}_i f_I{}^j =\delta_i^j \qquad & \qquad f^{I\, i}f_{I\, j}
=\delta_j^i\cr f^I{}_i f_I{}_j = 0 \qquad & \qquad f^{I\, i}f_I{}^j =0}
\eqn\sevend
$$
and
$$
f^I{}_i f_J{}^i + f^{I\, i}f_{J\, i} = \delta^I_J\ ,
\eqn\sevene
$$
from which it follows that
$$
f^I{}_i g_{IJ} = f_{J\, i}\ .
\eqn\sevenf
$$
These relations are invariant under local $U(n)$ transformations.
To preserve this invariance in the fermion couplings of the sigma
model we  must introduce an (anti-hermitian) $U(n)$ connection
one-form  $L{}^i{}_j$ which is determined in terms
of the frame forms by the requirement that
$$
\eqalign{
df^i + L^i{}_j f^j &=0\cr
df_i +f_j L^j{}_i &=0\ .}
\eqn\seveng
$$
It follows that
$$
F^i{}_j f^j = f_iF^i{}_j =0
\eqn\bianchi
$$
where
$$
F^i{}_j = dL^i{}_j + L^i{}_k L^k{}_j
\eqn\strength
$$
is the curvature two-form. As a consequence of \bianchi, it has the
property  that the tensor
$$
M_{ki}{}^{\ell j} \equiv f^{I\,\ell}f^J{}_k \, F_{IJ}{}^j{}_i
\eqn\symmetry
$$
is symmetric on its upper and lower indices.

The action for the N=2 supersymmetric sigma model in 2+1 dimensional
Minkowski  spacetime is
$$
\eqalign{
S= \int\! d^3 x\Big\{ {1\over2} &g_{IJ}\partial_\mu\phi^I\partial^\mu\phi^J
-\bar\lambda_i\gamma^\mu\big( \partial_\mu\lambda^i +\partial_\mu\phi^I
L_I{}^i{}_j\lambda^j\big)\cr
&\qquad -{1\over6}
M_{ij}{}^{k\ell}(\bar\lambda_k\gamma^\mu\lambda^i)(\bar\lambda_\ell\gamma_\mu
\lambda^j)\Big\} \
,}
\eqn\actionflat
$$
where the conjugate spinor $\bar\lambda_i$ is defined by
$$
\bar\lambda_i \equiv (\lambda^i)^\dagger \, i\gamma^\uzero\ .
\eqn\congdef
$$
This action is invariant under the supersymmetry transformations
$$
\eqalign{
\delta\phi^I &= {i\over2}f^I{}_i\bar\epsilon\lambda^i \ +c.c. \cr
\delta\lambda^i &= -{i\over2} f_I{}^i\gamma^\mu\epsilon\partial_\mu\phi^I -
L_I{}^i{}_j\delta\phi^I\lambda^j\ . }
\eqn\transflat
$$

We now turn to the construction of the action for the $N=2$
sigma-model  coupled to (2,0) adS supergravity. We shall use the
gamma-matrix  and other conventions of [1]. Note that because of our
$(+ - -)$  metric convention the gamma-matrices are all pure
imaginary. Also
$$
e\gamma^{\mu\nu\rho}=i\varepsilon^{\mu\nu\rho}\ ,
\eqn\iseveni
$$
where $e=\det e_\mu{}^a$ and
$\varepsilon^{\mu\nu\rho}$ is a (constant) tensor density. Note also that the
constant Lorentz-invariant tensor $\epsilon^{abc}$ is defined by
$$
\epsilon^{abc} = e^{-1} e_\mu{}^a e_\nu{}^b
e_\rho{}^c\varepsilon^{\mu\nu\rho}\ .
\eqn\sevenj\ .
$$

The fact that the two-form $\Omega$ is closed has the
consequence that the vector density
$$
j^\mu(\Omega)
={1\over2}\varepsilon^{\mu\nu\rho}\partial_\nu\phi^I\partial_\rho\phi^J
\Omega_{IJ}
\eqn\sevenk
$$
is identically conserved ($\partial_\mu j^\mu(\Omega)\equiv 0$). It is
therefore
a potential source for the Chern-Simons gauge field $A$ and the analogy with
supergravity/sigma-model actions in 4+1 dimensions suggests that such a term be
included. Indeed, the inclusion of such a coupling leads to the following
locally-supersymmetric and gauge-invariant action:

$$
\eqalign{
S &=\int d^3x\, \big\{
[-{1\over2}eR +{i\over2}\varepsilon^{\mu\nu\rho}\bar\psi_\mu
{\cal D}_\nu \psi_\rho
 - 2m\varepsilon^{\mu\nu\rho}A_\mu\partial_\nu
A_\rho +4m^2 e]\cr
&+{1\over2} e g_{IJ}\partial_\mu\phi^I\partial^\mu\phi^J
- A_\mu j^\mu(\Omega) - {1\over 8m}X_K\partial_\mu\phi^K j^\mu(\Omega)\cr
&-e \bar\lambda_i\gamma^\mu (\nabla_\mu
\lambda)^i  -{1\over2} e \big[ if_I{}^i\bar\lambda_i\gamma^\mu\dslash
\phi^I \psi_\mu
+ \ c.c.\big] - ime\, \bar\lambda_i\lambda^i\cr
& +{1\over8}
e\bar\lambda_i\lambda^i(\bar\psi_\nu\gamma^\mu\gamma^\nu\psi_\mu )
-{1\over8}e\bar\lambda_i\gamma^\rho\lambda^i\big(
\bar\psi_\nu\gamma^\mu\gamma_\rho\gamma^\nu\psi_\mu -\bar\psi\cdot
\gamma\psi_\rho
-\bar\psi_\rho\gamma\cdot\psi\Big)\cr
&-{1\over6}e
M_{ij}{}^{k\ell}(\bar\lambda_k\gamma^\mu\lambda^i)(\bar\lambda_\ell\gamma_\mu
\lambda^j)
-{1\over4}e(\bar\lambda_i\lambda^i)^2\  \big\}\ ,}
\eqn\sevenl
$$
where the covariant derivative $\nabla$ is defined by
$$
(\nabla_\mu \lambda)^i = D_\mu\lambda^i + \partial_\mu\phi^I
L_I{}^i{}_j \lambda^j + 2imA_\mu\lambda^i \ ,
\eqn\sevenm
$$
and the spin-connection used to define the Lorentz-covariant derivative $D$ is
now taken to be the one for which
$$
D_{[\mu}e_{\nu]}{}^a = {1\over4}\bar\psi_\mu\gamma^a\psi_\nu
 -{i\over4}e_{\mu b}e_{\nu
c}\epsilon^{abc}\bar\lambda_i\lambda^i\ .
\eqn\sevenn
$$
That is, the torsion now includes a contribution from the sigma-model fermions.

This action is invariant under the following supersymmetry transformations:
$$
\eqalign{
\delta_\epsilon e_\mu{}^a &= {1\over4}\bar\epsilon\gamma^a\psi_\mu
+ c.c.\cr
\delta_\epsilon\psi_\mu &= {\cal D}_\mu\epsilon -{1\over 4}\bar\lambda_i
\gamma^\nu\lambda^i \gamma_{\mu\nu}\epsilon\cr
\delta_\epsilon A_\mu &= {1\over4} \left( \bar\epsilon \psi_\mu
+c.c.\right)  -{1\over
4m}\delta_\epsilon\phi^I\partial_\mu\phi^J \Omega_{IJ}\cr
\delta_\epsilon\phi^I &= {i\over2}f^I{}_i\bar\epsilon\lambda^i + c.c.\cr
\delta_\epsilon\lambda^i &=
-{i\over2}f_I{}^i\hat\dslash\!\phi^I\epsilon -
L_J{}^i{}_j\delta_\epsilon\phi^J\lambda^j}
\eqn\seveno
$$
where
$$
\hat \partial_\mu\phi^I = \partial_\mu\phi^I
-[{i\over2}f^I{}_i\bar\psi_\mu\lambda^i
\ +\  c.c.]\ .
\eqn\sevenp
$$
It is also invariant under the gauge transformations
$$
\eqalign{
\delta_\Lambda \psi_\mu =-i\Lambda\psi_\mu\qquad &\qquad
\delta_\Lambda A_\mu = (2m)^{-1}\partial_\mu\Lambda\cr
\delta_\Lambda\phi^I =0 \qquad &\qquad
\delta_\Lambda\lambda^i = -i\Lambda \lambda^i \ ,}
\eqn\sevenq
$$
for arbitrary function $\Lambda(x)$.
%%%%%%%%%%%%%%%%%%%%%%%%%%%%%%%Chapter6%%%%%%%%%%%%%%%%%%%%%%%%%%%%%%%%

\chapter{ Supersymmetric self-gravitating solitons}

For present purposes we need only the
bosonic sector of the supergravity coupled sigma model action just
constructed. This is
$$
\eqalign{
S_{bos.}=\int d^3x\, \big\{ -{1\over2}& eR -
2m\varepsilon^{\mu\nu\rho}A_\mu\partial_\nu A_\rho +4m^2e
+{1\over2} e g_{IJ}\partial_\mu\phi^I\partial^\mu\phi^J \cr
&- A_\mu j^\mu(\Omega) -{1\over 8m}X_K\partial_\mu\phi^K j^\mu(\Omega)
\big\}\ . }
\eqn\eighta
$$
We shall also need the fermion supersymmetry transformations in a purely
bosonic background. These are
$$
\eqalign{
\delta_\epsilon\psi_\mu &= {\cal D}_\mu\epsilon\equiv (\partial_\mu +
iB_\mu{}^a\gamma_a +
2imA_\mu )\epsilon\cr
\delta_\epsilon\lambda^i &= -{i\over2}f_I{}^i\dslash\phi^I\epsilon\ ,}
\eqn\eightb
$$
where, since the torsion vanishes in a purely bosonic background,
$B_\mu{}^a$  is given by \threee.

The curious final `topological' term in the action
\eighta\ deserves comment. For the special case of $\cM =S^2\,$ it is
just the  well-known Hopf term [\WZ]. Its presence is required by
supersymmetry if the coupling of $A_\mu$ to the topological current
$j^\mu(\Omega)$ is included but its coefficient is such that the field
redefinition $$ A'_\mu = A_\mu + {1\over4m}X_K\partial_\mu \phi^K
\eqn\eighte
$$
leads to the new, and much simpler, action
$$
S_b=\int d^3x\, \big\{ -{1\over2} eR -
2m\varepsilon^{\mu\nu\rho}A'_\mu\partial_\nu A'_\rho +4m^2e
+{1\over2} e g_{IJ}\partial_\mu\phi^I\partial^\mu\phi^J \big\}
\eqn\eightf
$$
in which both the Hopf-type term and the coupling of $A$ to the
topological  current density $j(\Omega)$ are absent. However, this redefinition
would complicate the supersymmetry transformations. Written in terms
of $A$  (rather than $A'$) the supersymmetry transformation of
$\psi_\mu$ is  exactly as it was in the absence of
matter, apart from $\lambda^2$ terms which vanish in a purely bosonic
background. This has the consequence that the Killing spinor equation
found  previously in the context of the pure (2,0)
adS supergravity theory is {\it unchanged by the coupling to matter}.
This is  a special feature of the particular model that we have
constructed;  more general matter couplings exist but they
will not be discussed here.

The field equations of the action \eighta\ are
$$
\eqalign{
S^\mu{}_a &=-{1\over2}T^\mu{}_a \cr
G^\mu &= {1\over 2m} J^\mu }
\eqn\eightc
$$
and
$$
\nabla_\mu \partial^\mu\phi^J =0\ ,
\eqn\eightd
$$
where
$$
\eqalign{
T_{\mu\nu} &= g_{IJ}\Big( \partial_\mu\phi^I\partial_\nu\phi^J
-{1\over2} g_{\mu\nu}
\partial_\rho\phi^I\partial^\rho\phi^J\Big)\cr
J^\mu &= -j^\mu(\Omega)\ , }
\eqn\currents
$$
and $\nabla$ is the standard spacetime and target space covariant derivative.
The field equations \eightc\ are precisely of the form assumed earlier
in our  derivation of energy bounds. From the specific form of the
currents  one may verify that
$$
\gamma^\nu T^\mu{}_\nu -J^\mu = g^{IJ}\Big(\bar
P_{IK}\dslash\!\phi^K\Big)  \gamma^\mu
\Big( P_{JL}\dslash\!\phi^L\Big)\ ,
\eqn\iden
$$
where
$$
P_{IJ} = {1\over2}\Big( g_{IJ} +i\Omega_{IJ}\Big)
\eqn\proj
$$
and $\bar P$ is its complex conjugate. It follows, for any spinor $\chi$, that
$$
-i\bar\chi \Big(\gamma^\nu T^\mu{}_\nu -J^\mu\Big)\chi =
-ig^{IJ}\bar\Psi_I\gamma^\mu\Psi_J\ ,
\eqn\encon
$$
where $\Psi_I\equiv P_{IJ}\dslash\!\phi^L\chi$. The right hand side of
\encon\  is manifestly non-spacelike and future directed, so the
condition on  the stress tensor and $U(1)$ current required to
establish the  classical energy bound in section 3 is indeed satisfied by
supersymmetric sigma-model matter. As we saw in that section the
energy bound  can be saturated only if $\chi=\kappa$ where $\kappa$ is
a  Killing spinor. It is also clear from \sixb\ and \encon\ that a
further  condition for saturation of the bound in the presence of sigma model
matter is that, when $\chi=\kappa$, $\psi_I=0$. Not surprisingly, this
is equivalent to the condition that $\delta_\epsilon \lambda^i=0$ if
the  anti-commuting complex spinor parameter $\epsilon$ is replaced by
the  complex commuting Killing spinor $\kappa$. Thus, to find
supersymmetric non-vacuum spacetimes we must find sigma model fields
$\phi^I$  such that
$$
P_{IJ}\dslash\phi^J\kappa =0 \eqn\eightg
$$
where $\kappa$ is a
Killing  spinor. Given that $\kappa$ has the form $\kappa
=(1+b_a\gamma^a)\zeta$, where $b_a$ are the functions introduced
in section 2, we find that \eightg\ is equivalent to
$$
{1\over2}(g_{IJ}+i\Omega_{IJ})\partial_\mu\phi^J(e_c{}^\mu +
i\varepsilon^{ab}{}_c\, e_a{}^\mu
b_b)=0\ . \eqn\eighth
$$

To solve these equations we consider matter fields of the form
$\phi^I(\varphi  +2mt,r)$ and we try the ansatz
$$
\partial_\varphi\phi^I = \gamma(r)\Omega^I{}_J\partial_r\phi^J\ .
\eqn\eighti
$$
This leads to a solution of \eighth\ provided that
$$
\gamma = {rf\over h\beta}\ ,
\eqn\eightj
$$
where $\beta$ is the constant introduced in section 2, and the
function $f$ is  given by \foure\ (with $\alpha=2m$). Clearly we must
exclude  $\beta=0$ here but it follows from \fouri\
that $G^\mu=0$ when $\beta=0$, and then from \eightc\ and \currents\
that  $\phi^I$ must be constant, so no generality is lost by this exclusion.

We have now reduced the solution of the conditions \eighth\ for the
matter fields to preserve supersymmetry to the problem of finding functions
$\phi^I(\varphi +2mt, r)$ satisfying \eighti. If we introduce the complex
spacetime coordinate
$$
z= exp[ \int\!\gamma^{-1}dr + i(\varphi +2mt)]\ ,
\eqn\eightl
$$
and choose complex coordinates $\phi^\alpha$ ($\alpha=1,\dots,n$) on the target
space, then the complex fields $\phi^\alpha(\varphi +2mt,r)$ become
functions  of $z$ and $\bar z$ and \eighti\ reduces to
$$
{\partial\phi^\alpha \over \partial \bar z}=0\ .
\eqn\eightm
$$
which is solved by {\it holomorphic} functions of $z$. This is
precisely the  condition found for sigma-model solitons in flat space.

The equation of motion \eightd\ for $\phi^I$ is now found to be
identically  satisfied so the holomorphic functions $\phi^\alpha(z)$,
as well  as the function $u$, must be found from the
equations \eightc\ which, given the results of section 2, are found to
reduce  to the single equation
$$
{\beta^2\over 2rf^3h} (fh)' =
g_{IJ}\partial_r\phi^I\partial_r\phi^J\ .
\eqn\eightn
$$
Using complex coordinates on the target space, and the fact that the fields
$\phi^\alpha$ are holomorphic functions of $z$, we can rewrite this equation as
$$
{r\over 2h^3f} (fh)' = g_{\alpha\bar\beta}|z|^2 \partial_z\phi^\alpha
\partial_{\bar
z}\bar\phi^{\bar\beta}\ .
\eqn\eighto
$$
Since the functions $f$ and $h$ are determined in terms of $u$ by \foure, this
equation can be viewed as a second order ODE for $u$ with a source
determined  by the holomorphic functions $\phi^\alpha$. However, the
LHS of  \eighto\ depends only on $r$, i.e. on $|z|$, while
this is true of the RHS only for very special holomorphic functions.
For  simplicity, we shall now restrict the discussion to the simplest
possible  K{\" a}hler target manifold, the Riemann sphere,
parametrized by a single complex coordinate $\phi$, in which case
\eighto\  becomes
$$
{r\over
2h^3f} (fh)' = {4\over (1+|\phi|^2)^2}|z\partial_z \phi|^2\ .
\eqn\eightp
$$
In order that the right hand side be a function of $|z|$ only we are
forced to  set
$$
\phi = Cz^k
\eqn\eightk
$$
for some constant $C$ and integer $k$. The simplest case is $k=0$,
i.e.  constant matter fields.
The first non-trivial case is $k=1$ and we shall examine this case in
some  detail. Setting $\phi=Cz$ in \eightp\ we have
$$
{r\over 2h^3f} (fh)' = {4|Cz|^2  \over (1+|Cz|^2)^2}\ .
\eqn\eightl
$$
To proceed we use the fact that
$$
{\partial\over\partial r} = {\beta h|z|\over rf}{\partial\over\partial |z|}
\eqn\eightm
$$
to rewrite \eightl\ as an ODE with $|z|$ as the independent variable.
This ODE  can be immediately once-integrated to give
$$
{4\over 1+|Cz|^2} = c +{\beta\over fh}\ ,
\eqn\eightn
$$
where $c$ is the integration constant. We note for future reference
that $fh$  can be constant only if $C=0$, which corresponds to
vanishing  matter, or $C=\infty$, which is clearly unphysical.
Constant $fh$ implies constant $u$, i.e a vacuum solution (as expected
for  $C=0$) and thus $fh=1$
by our analysis of section 2. It follows that
$$
\eqalign{
C=0 &\Rightarrow c= 4-\beta\cr
C=\infty &\Rightarrow c= -\beta\ .}
\eqn\aeightn
$$

Using \eightn\ to eliminate $|Cz|$ from \eightl\ we obtain the ODE
$$
rf^2 (fh)' =  -{1\over2} c(c-4)(fh)^3 -\beta (c-2) (fh)^2 -{\beta^2\over2}
(fh)\ .
\eqn\eighto
$$
Defining the new dependent and independent variables
$$
r^2=y\qquad \tilde u(y) = 2mr^2 -u(r) \ ,
\eqn\eightp
$$
this equation can be simplified to
$$
(\tilde u^2 + \beta^2 y)\tilde u''= -{c(c-4)\over 16m^2} (\tilde u')^3 -
{\beta (c-2)\over 4m}(\tilde u')^2 -{\beta^2\over4} \tilde u'
\eqn\eightq
$$
We have not been able to find an analytic solution of this second
order  non-linear differential equation, but assuming that $\tilde u
\sim   2mr^2$, asymptotically as $r\rightarrow \infty$ the asymptotic
solution  for $u$ takes the form
$$
u\sim -{K\over 4m}\ln r + {1\over2}\tilde J + {\cal O}\Big({\ln r\over
r^2}  \Big)\qquad
(r\rightarrow\infty)
\eqn\asymp
$$
where $\tilde J$ is a constant and
$$
K= (c+\beta)(c+\beta -4)
\eqn\normlog
$$
is another constant. Because of the logarithmic term, the constant
$\tilde J$  cannot be identified with the total angular momentum
unless $K=0$,  hence the change in notation. For
convenience we recall that the metric under discussion takes the form
$$
ds^2 = \big((2mr)^2 -4m u + \beta^2\big)dt^2 -2u dtd\varphi
-{\big[1 -\left({u'\over 4mr}\right)\big]^2\over
\big[\big(2m r - \left({u\over r}\right)\big)^2 + \beta^2\big]} dr^2
-r^2d\varphi^2\ .
\eqn\fourlagain
$$
 From \asymp\ we see that its asymptotic form is
$$
\eqalign{
ds^2 \sim \Big[ &(2mr)^2 + K \ln r - 2m\tilde J + \beta^2 +
{\cal O}\Big({\ln r\over r^2} \Big)\Big] dt^2 \cr
& - \Big\{ \Big[ (2mr)^2 + K \ln r -2m\tilde J + \beta^2 \Big]^{-1} +
{\cal O}\Big({1\over r^2} \Big)\Big\} -r^2 d\varphi^2 \ .}
\eqn\asmetric
$$
The corresponding expressions for the asymptotic behaviour of the
fields $A$  and $\phi$ are
given by
$$
\eqalign{
A &\sim \Big[ {\beta K\over 32m^2 r^2}
+ {\cal O}\Big({\ln r \over r^4}\Big)\Big] dt \cr
&\qquad + \Big[-{(\beta +n)\over 4m} - {\beta K\over 64m^3 r^2} +
{\cal O}\Big({\ln r\over r^4}\Big)\Big]d\varphi\cr
\phi &\sim C e^{i(\varphi +2mt)}\Big[ 1-{\beta\over 2mr} +
{\cal O}\Big({1\over r^2}\Big)\Big]\ .}
\eqn\extraone
$$
The logarithmic term in \asymp\ vanishes when $K=0$, i.e. when either
$c=-\beta$ or $c=4-\beta$. In fact, all terms in $u$ other than the
constant  vanish when $c$ takes one these two values,
which therefore lead to vacuum solutions. Since $fh=1$ for a vacuum
solution  (given $\alpha=2m$), we learn that the implications of
\aeightn\ can be reversed. That is, $c=4-\beta$ implies that
$C=0$, which we now exclude since we are interested in soliton
solutions, and  $c=-\beta$ implies that $C=\infty$, which we exclude
as  unphysical. We conclude that the logarithmic term is {\it
necessarily present} for self-gravitating solitons, so $K\ne 0$. When
$\tilde  J=0$ the $K=0$ case can be thought of as the limiting case in
which  the soliton becomes a point particle at the origin, which is
then a  conical singularity. It is not difficult to see that the
logarithmic  term is a {\it necessary} feature for a spacetime to be
non-singular without horizons and we argue below that the spacetimes
of the  self-gravitating solitons found here indeed have this property.

 From \extraone\ we see that the vector potential $A$ tends to a constant as
$r\rightarrow \infty$ and the complex scalar field $\phi$ tends to a
phase.  From this it is easy to see that both the field strength
two-form for  $A$ and the energy momentum tensor for $\phi$
vanish at infinity. But since $u$ grows logarithmically the metrics of
our  soliton solutions are not asymptotically anti de Sitter in the
sense of  [\BH] so the results of those authors is not
applicable. Nor is it clear that the charges defined as in section 4
make  sense. However, although neither $M$ nor $J$ is separately
well-defined,  the linear combination $\tilde M =M-2mJ$ {\it is},
because the  potentially divergent contribution due to the $\ln r$ term in $u$
cancels, and we may as well define this quantity to be the soliton's
mass; it  saturates the bound $\tilde M\geq -\beta^2$, as in the vacuum case.

Singularities of the metric \fourl\ for $u$ given by \eightp\ and
\eightq\ can  arise, in principle, either from some component becoming
infinite at finite $r$ or because the inverse fails to exist. Any such
singularity may be merely a coordinate singularity, of course; this
point must be adressed after location of the singularities. A
component of the  metric can become infinite for finite $r$ only if
$u$ becomes infinite for finite $r$, or when $u$ is of the form
$u\sim r$ as $r\rightarrow 0$, in which case $g_{rr}$ would become infinite.
As we shall see, neither case can occur. Consider first the case
$u\rightarrow\infty$ as $r\rightarrow r_0$ or, equivalently, $\tilde
u\rightarrow \infty$ as $y\rightarrow y_0=r_0^2$. In
this case it is convenient to reinterpret \eightq\ as an equation for
$y$ in  terms of $\tilde u$, in which case \eightq\ can be rewritten
as
$$  (\tilde u^2 +\beta^2y) y'' = {c(c-4)\over 16m^2}
+{\beta(c-2)\over 4m} y'  +{\beta^2\over 4} (y')^2 \eqn\extratwo
$$
where now $y= y(\tilde u)$
and $y'= dy/d\tilde u$. We have to look for solutions $y$ of this
equation  that go to a constant as $\tilde u$ goes to infinity. Let us
write  the asymptotic behaviour of $y$ as
$$
y= y_0 +
y_1(\tilde u) \eqn\extrathree
$$
where $y_1$ goes to zero as $\tilde u$ goes to infinity.
Neglecting non-leading terms we have (unless $c=0$ or $c=4$) that
$$
(\tilde u)^2 y_1'' \approx
{c(c-4)\over 16m^2} \eqn\extrafour
$$
the solution of which is $y_1 \propto \ln\tilde u$, contrary to
assumption.  For the cases $c=0$  or $c=4$ we instead find the equation
$$
(\tilde u)^2 y_1'' \approx y_1'\ ,
\eqn\extrafive
$$
the solution of which is again inconsistent with the asymptotic
condition for  $y_1$. We conclude that $u$ is finite for finite $r$.
To  exclude the other possibility, that $u\sim r$ as $r\rightarrow 0$,
we  shall suppose that  $\tilde u \sim y^\alpha$, for some constant
$\alpha$,  as $y \rightarrow 0$. One then discovers that this is
consistent  with \eightq\ for $\alpha=0$ or
$\alpha=1$, but not otherwise.

It remains to determine whether the metric is invertible. Since
$$
\det g = r^2 (fh)
\eqn\extraeight
$$
the metric will be singular, apart from the expected singularity at
the origin  $r=0$ of the polar coordinates, only if $(fh)$ vanishes.
It  follows from \eightl\ that $(fh)\rightarrow -\infty$ as
$r\rightarrow  \infty$ if $(fh)$ is anywhere negative. As this
contradicts the  known behaviour $(fh)\rightarrow 1$ as $r\rightarrow
\infty$,  we conclude that $(fh)$ cannot be negative. Furthermore,
\eightl\  implies that $(fh)'$ must vanish at any point, other than $r=0$,
at which $(fh)$ vanishes. Observe now that linearization of \eighto\
about a  point at which $(fh)$ vanishes leads to the conclusion that
$(fh)'$  is everywhere negative in some neighbourhood of this point,
but this  is not possible unless $(fh)$ changes sign there. Since it
cannot  change sign our supposition that there was a point at which
$(fh)=0$  must be false. Thus $r=0$ is the
only point at which the metric is singular.
As we have seen, the only possible power behaviour of $\hat u$ near
the origin  is $\hat u\sim y^\alpha$ for $\alpha=0$ or $\alpha=1$. In
the  latter case \eightq\ also determines the coefficient, such that
$\tilde u  = ay + o(y)$, and $a=-{2m\over c-4}\beta$ (the value
$a=-{2m\over c}\beta$ is also allowed by the equation but we exclude
it since  $c=4-\beta$ should correspond to $a=2m$). If we require the
singularity of the metric at $r=0$ to be a
mere coordinate singularity we further discover that the constant $c$
must be chosen such that $\tilde u = 2m\beta y + O(y^2)$ (note that
this  includes adS for $\beta=\pm 1$). When $\tilde u$ tends to a
non-zero  constant as $r\rightarrow 0$, non-singularity at $r=0$
restricts the  next to leading term so that $u\propto 1 \pm 4mr +
O(r^2)$, athough this possibility is excluded by the requirement that
$A$ and  $\phi$ also be non-singular at $r=0$, as we shall see. We
conclude  that \eightq\ is compatible with a
non-singular metric at the origin.

We now turn to the behaviour of the other fields near the origin. From
\fouri\  we see that non-singularity of $A$ requires that $dA$ vanish
at  $r=0$. This restricts the value of $n$: to $n=0$ when $\tilde u
\sim  const.$ as $r\rightarrow 0$ and to $n=-1$ when $\tilde u \sim 2m\beta
y$ as $\rightarrow 0$. However, when $\tilde u\sim const.$,
$|\phi|\sim  const.$ as $r\rightarrow 0$, which (since the constant is
non-zero) implies that $\phi$ is singular at $r=0$. On the other
hand, when  $\tilde u \sim 2m\beta y$, $\phi|\sim r$ as $r\rightarrow
0$, so  $\phi$ is non-singular at $r=0$. Thus, non-singular sigma
model  matter requires $|n|=1$. This is to be expected because when
$u\sim r^2$ as $r\rightarrow 0$, the Killing spinor $\kappa$ tends to a
constant and this is compatible with its non-singularity at $r=0$ only
for  $|n|=1$.

Thus, the behaviour near $r=0$ of the metric, the CS gauge field, and
the  sigma-model scalars, is compatible with the existence of a
non-singular  soliton solution admitting Killing spinors
with $|n|=1$. Near $r=0$ the metric is of the form \fourlagain\ with
$$
u(r) \sim 2m(1-\beta)r^2 + {\cal O}(r^4) \qquad (r\rightarrow 0)\ .
\eqn\unearzero
$$
To complete the proof that non-singular self-gravitating solitons
exists we  would need to show that the solutions of \eightq\ that
behave like  \unearzero\ near the origin match onto solutions
with the asymptotic behaviour \asymp\ as $r\rightarrow \infty$. The
non-linearity of the equation for $u$ makes this a difficult problem
which we  have not solved.

%%%%%%%%%%%%%%%%%%%%%%%%%%%%%%%Chapter 7%%%%%%%%%%%%%%%%%%%%%%%%%%%%%%%%

\chapter{ Conclusions}

The work reported here was initially motivated by some similarities
between  2+1 and 4+1 dimensions. However, it is ultimately the
differences  that are most striking. One of these is the
fact that there is not just one energy bound but at least two. There
are, in  fact an infinite number of supersymmetric static spacetimes
with  $M=-n^2$, so that supersymmetry alone does not
imply a lower bound on the energy, although only two of these (adS
spacetime  and the black hole vacuum) have non-singular Cauchy
surfaces; the  spacetimes for $n>1$ all have a naked conical
singularity with negative deficit angle. These facts are related to
the fact  that, like asymptotically Kaluza-Klein spacetimes, spatial
infinity  is multiply connected. For this reason one must distinguish
between  periodic and antiperiodic spinors when deriving the BGH bound.
Assuming that this distinction is sufficient for spacetimes that are
non-singular on some initial Cauchy surface and that solve the
Einstein  equations, with matter stress tensor satisfying the
dominant energy condition, we concluded that the adS spacetime is
absolutely  stable, quantum mechanically, as well as classically but
that the  black hole vacuum might be unstable against
quantum tunnelling. This deserves futher investigation.

Yet another surprise is that not only do the $M=2m|J|$ extreme black
hole  spacetimes admit Killing spinors, as shown in [\HC], but so also
do the  spacetimes with $M=-2m|J|$, although again
these negative mass spacetimes have naked cusp-like singularities and
are  presumably unphysical. The spacetimes associated with point
particles  [\DJ] with a mass in the range $-1<M<0$ have a
naked conical singularity at the origin with deficit angle $\delta=
2\pi(1+\sqrt{M})$. We have shown that these spacetimes are
supersymmetric if  the particle has a charge $Q$ such that
$M=-4Q^2$ or, equivalently
$$ E> 1-4Q^2 \ ,
\eqn\bog
$$
where $E$ is the energy with respect to the adS vacuum. We have also
shown  that this relation can be interpreted as the saturation of a
Bogomolnyi-type bound by a supersymmetric solution of
the field equations.

The supersymmetric self-gravitating solitons that we have found can be
considered as smeared versions of these supersymmetric point
particles,  although the asymptotic behaviour is quite different. They
share some features with self-gravitating instantonic solitons of
five-dimensional supergravity. In the latter case, the curved space
equations  for the Yang-Mills matter reduced to the flat space
self-duality equations as a result of particular interactions of
the Maxwell field that were required by supersymmetry [\GKLTT]. In the
2+1  case we similarly found that the holomorphicity condition for
supersymmetric flat space sigma-model solitons is
maintained by the coupling to adS supergravity, but there was a
further  restrictive condition. In fact, the self-gravitating
sigma-model  solitons are not strictly asymptotic to anti de Sitter
spacetime because of a logarithmic terms in the metric. The criteria
for a  solution of adS gravity to be a {\it bone fide}
self-gravitating  soliton clearly deserves further study. One
wonders, for example, whether the solitons we have found could be
pair-produced. The same question could reasonably be asked of the
extreme  black holes.

Finally, those new supersymmetric spacetimes found here for which
$Q=0$ can  also be considered as solutions of the effective field
equations  of three-dimensional string theory [\HW], so that they will
have  target space duals associated with exactly the same string
theory. It will be of interest to see whether any of the singular
supersymmetric asymptotically adS spacetimes have non-singular duals.

%%%%%%%%%%%%%%%%%%%%%%%%%%%%%%%%%%%%%%%%%%%%%%%%%%%%%%%%%%%%%%%%%%%%%%%

\noindent{\bf Acknowledgements:} Discussions with Gary Gibbons, Jim
Horne,  Paul Howe, George Papadopoulos and Alan Steif are gratefully
acknowledged.

\refout

\end